# Game-Theoretic Deep Reinforcement Learning to Minimize Carbon Emissions and Energy Costs for AI Inference Workloads in Geo-Distributed Data Centers

Ninad Hogade, *Member, IEEE,* and Sudeep Pasricha, *Fellow, IEEE*

**Abstract** — Data centers are increasingly using more energy due to the rise in Artificial Intelligence (AI) workloads, which negatively impacts the environment and raises operational costs. Reducing operating expenses and carbon emissions while maintaining performance in data centers is a challenging problem. This work introduces a unique approach combining Game Theory (GT) and Deep Reinforcement Learning (DRL) for optimizing the distribution of AI inference workloads in geo-distributed data centers to reduce carbon emissions and cloud operating (energy + data transfer) costs. The proposed technique integrates the principles of non-cooperative Game Theory into a DRL framework, enabling data centers to make intelligent decisions regarding workload allocation while considering the heterogeneity of hardware resources, the dynamic nature of electricity prices, inter-data center data transfer costs, and carbon footprints. We conducted extensive experiments comparing our game-theoretic DRL (GT-DRL) approach with current DRL-based and other optimization techniques. The results demonstrate that our strategy outperforms the state-of-the-art in reducing carbon emissions and minimizing cloud operating costs without compromising computational performance. This work has significant implications for achieving sustainability and cost-efficiency in data centers handling AI inference workloads across diverse geographic locations.

—————————— ◆ ——————————

## 1 INTRODUCTION

The internet has evolved into a crucial aspect of modern life, with billions of users worldwide and the ability to support various services and apps. Approximately 5.25 billion people have access to and utilize the internet globally as of 2023, making up 66.2% of the expected 7.9 billion people on the planet [1]. The use of data centers to support Internet applications is growing rapidly, with experts predicting increased reliance on data centers fueled by Artificial Intelligence (AI)-driven workloads, new cloud services, demand for edge applications, expansion of the Internet of Things (IoT) devices, and the proliferation of technologies such as 5G/6G.

The rise of AI workloads in data centers has been a significant trend in recent years. AI workloads require immense computing power, especially for training complex deep learning models. They also necessitate vast amounts of data for training and improving models, leading to increased data transfer and storage needs. As a result, AI-driven computing is fueling demand for more power-hungry data centers with expanded storage capabilities [2]. In addition to storage and computing power, the widespread use of AI workloads has led to a growing need for real-time processing and decision-making. These workloads have gained prominence in data centers due to several factors, including the growth in smart edge computing (e.g., with increasingly autonomous vehicles) and the proliferation of IoT applications (e.g., those used in smart home devices) [3].

Cloud service providers are investing significantly in extending their data center infrastructure to accommodate the rising demand for cloud services and AI inference workloads. In particular, there has been a shift towards building and adding data centers across multiple geographic locations [9], [10]. There are several reasons to favor geographically distributed data centers as they provide advantages in load balancing, reliability and redundancy, latency reduction, and compliance with data sovereignty rules [11]. Another compelling reason for geographically spreading data centers is to reduce electricity costs by taking advantage of Time-Of-Use (TOU) electricity pricing. According to the TOU electricity pricing model [12], electricity prices fluctuate depending on the time of day. When demand for the entire electrical grid is high, electricity costs rise; whereas when the demand is low, they decline [13]. Utilities also impose an additional flat-rate (peak demand) cost on non-residential or commercial customers depending on the maximum (peak) power utilized at any given moment within a monthly billing period [14]. These peak demand fees sometimes exceed the amount of the electricity bill devoted to energy use. One efficient way to lower electricity costs is to move workloads between geo-distributed data centers. With the help of this strategy, cloud service providers can distribute workloads to data centers with lower energy prices. This can reduce the cost of cloud computing for clients and operating costs for cloud service providers.

Today, major cloud service providers run multiple massive power-hungry data centers around multiple geographic locations. These data centers consume about 3% of global electricity and contribute approximately 2% of all GreenHouse Gas emissions (GHG) emissions worldwide [15]. During the COVID pandemic, increased reliance on web and

- *N. Hogade is with Hewlett Packard Labs, Fort Collins, CO 80528. E-mail: ninadhogade@gmail.com*
- *S. Pasricha is with the Dept. of Electrical and Computer Engineering, Colorado State University, Fort Collins, CO 80523. E-mail: sudeep@colostate.edu.*

cloud services drastically increased internet traffic and data center utilization. If this trend continues, it has been predicted that data center GHG emissions may increase to 5–7% of global emissions, which is a major concern [15]. Thus, making data centers carbon-conscious and energy efficient has become a top priority for operators and cloud service providers.

When making decisions concerning AI inference workload management in geographically dispersed data centers, it is crucial to consider all the factors mentioned above. It is a challenging (NP-hard) problem to orchestrate complex, heterogeneous, large-scale software and hardware systems and automate computing platforms used in geo-distributed cloud computing systems. Researchers in this area have traditionally solved such problems using several mathematical methods, such as linear, non-linear, convex, integer, stochastic, combinational, dynamic, heuristic, probability, and hybrid optimization approaches. As discussed in our previous work [16] and [17], some algorithms are very good (e.g., genetic algorithms) at finding solutions but take too long to execute. At the same time, some algorithms are fast but susceptible to getting stuck in local minima (e.g., greedy and game-theoretic algorithms) [18], [17]. The real-time variability of workloads, combined with changing operating conditions such as data center availability, co-location interference, network costs, renewable power, electricity and peak demand pricing, needs a dynamic optimization strategy. Owing to the computational intensity and intricate characteristics of the data, coupled with the continuously changing problem space, such mathematical methods often exhibit limited applicability in large-scale dynamic distributed systems and encounter challenges in scalability concerning geographically distributed architectures. As a result, researchers have been investigating intelligent data-driven and Reinforcement Learning (RL) based optimization alternatives.

This work aims to create and assess a workload management strategy for geographically distributed data centers to reduce cloud carbon emissions and operational costs across all data centers for processing incoming workloads. This study acknowledges the correlation between energy costs and carbon emissions in data centers, but treats them as separate optimization objectives. We define cloud carbon emissions as the total carbon emissions of all geo-distributed data centers a cloud service provider uses. Similarly, we define cloud operating costs as the total operating (energy and network) costs of all combined geo-distributed data centers in the cloud infrastructure. We use real workloads and arrival patterns and compare our approach to state-of-the-art mathematical techniques and other DRL methods to evaluate it. Our work makes the following innovative contributions, in brief:

- we formulate the cloud workload distribution problem as a Nash equilibrium-based non-cooperative game;
- we then combine the game theory with deep reinforcement learning (DRL) and design a new intelligent game-theoretic DRL (GT-DRL) workload distribution framework to minimize the overall cloud carbon emissions and operating costs for geographically distributed heterogeneous data centers;
- the workload scheduling decisions leverage detailed models for a comprehensive set of characteristics that impact cloud carbon emissions and workload performance, including data center compute and cooling power, co-location performance interference, and variable renewable power available across different data center locations.

The remainder of this article is organized as follows. We discuss pertinent previous work in Section 2. In Section 3, we describe our system model. Our workload distribution problem is described in Section 4, while Section 5 presents the approach to solving the problem. Section 6 describes the simulation environment. In Section 7, we analyze and assess the results of our framework. Lastly, we conclude this article in Section 8.

## 2 BACKGROUND AND RELATED WORK
### 2.1 Data Centers and AI Inference

It is commonly known in the world of cloud data centers that CPU-based nodes outnumber GPU-based nodes, mainly because of their versatility and ability to handle a wide range of computational workloads [4]. The AI inference market is getting bigger than AI training [5]. This dominance is important because AI inference workloads are starting to take center stage and are requiring more computing power than AI training workloads [6]. While AI inference workloads are typically less computationally intensive than AI training workloads, they are nevertheless necessary for real-time applications like voice assistants and autonomous driving and require high efficiency and low latency [7]. CPUs play a significant role in AI inference due to their ability to do complicated statistical computations for natural language processing and deep learning algorithms. Advancements in technology, such as Intel's AMX units in Sapphire Rapids CPUs, have improved performance, making CPUs a viable solution for AI inference workloads [4]. The optimization of CPU infrastructure for AI inference is crucial for achieving optimal performance. Disabling hyperthreading can improve AI inference workload performance by optimizing resource allocation and lowering latency, leading to higher throughput and efficiency [7]. Furthermore, using the appropriate system profiles, such as the NFVI FP Energy-Balance Turbo Profile, can significantly improve AI inference performance on CPUs. This profile optimizes system performance for low-precision mathematical calculations in AI inference workloads while remaining energy efficient [7]. These strategic infrastructure improvements underscore the crucial need of adapting data center operations to suit the rising demands of AI inference, exploiting the vast availability and progressive capabilities of CPU-powered nodes. CPUs are more cost-effective and energy-efficient than GPUs when conducting inference tasks, making them a more sustainable alternative for extensive, large-scale AI inference deployments [8]. In light of these considerations, it is clear that CPU-based nodes provide a compelling, cost-effective, and energy-efficient solution for AI inference workloads, which is supported by ongoing research and technology improvements. That is why we focus on AI inference workloads and the CPU-based computing hardware in this work.

### 2.2 Reinforcement Learning

In RL, an agent learns to solve a problem by trial and error through interacting with the environment. It uses a feedback loop that includes rewards, which act as prompt assessments of the efficacy of the prior action. A reward is an immediate return (feedback) from the environment to assess the quality of the previous action. The agent represents the RL algorithm, while the environment refers to the item on which the agent is acting. The environment sends a current state to the agent to start the RL process. The agent chooses an action based on prior knowledge or a predetermined policy, where the policy is the agent's strategy for determining the next course of action based on the current state. The environment gives the agent the new state and related reward after completing an action. The agent then updates its knowledge or policy using this information. The cycle continues until the environment sends a terminal state, ending the episode. By combining Deep Learning (DL) with RL, Deep Reinforcement Learning (DRL) has been demonstrated to work better than traditional RL algorithms [19]. As discussed in our literature review on machine learning approaches for geo-distributed data center management [20], many works have recently explored using DRL, showing the feasibility of such solutions in real cloud environments. In DRL, Deep Neural Networks (DNNs) approximate the policy. To maximize its cumulative reward over time, the agent takes a sequence of actions and learns how to make better decisions by updating its DNN (policy).

## 2.3 Related Prior Work

We analyzed the recent literature on RL-based workload management for geo-distributed data centers. There are two primary types of RL algorithms: model-based and model-free RL. In model-based RL, the agent tries to create a model of the environment and optimizes its policy based on this model. In [21], the authors use a model-based RL as part of a two-step control loop in which a network-aware placement policy allows containers on geographically distributed computing resources, and a model-based RL technique dynamically adjusts the number of copies of individual containers based on the workload response time. Model-based RL techniques work well for a small problem where all environment variables are known. However, the solution space grows when the problem becomes large, such as in the case of the geo-distributed heterogeneous workload to data center resource mapping problem. In these cases, such algorithms become impractical.

On the other hand, model-free RL is independent of an environmental model. Instead, the agent uses a trial-and-error strategy to determine the best action. Model-free RL can be divided into two methodologies: Q-learning (value iteration) and policy optimization (policy iteration). Q-learning learns the action-value function. In other words, it aims to identify the best action for a particular state using a Q-table, a lookup table that stores the highest possible predicted rewards for actions at each state. The authors from [22], [23], [24], and [25] use Q-learning for their RL-based geo-distributed data center workload management. Deep Q-Network (DQN) is a variation of Q-learning that uses DNNs. DQN has been used for large problems with large state-action spaces where creating a Q-table would be extremely difficult, time-consuming, and complex. In DQN, DNNs approximate Q-values for each action based on the state. The approaches presented in [26], [27], [28], [29], [30], and [31] use DQN for run-time workload to resource mapping for geo-distributed data centers.

In policy optimization methods, the agent directly learns the policy function that maps the state to action. A value function or Q-table is not used to decide the policy. Actor-Critic methods combine both Q-learning and policy optimization. The critic component estimates the action-value (Q-value) or state-value function, while the actor adjusts the policy distribution according to the critic's suggestions. While some complex optimization problems can be solved using a discrete action space, others require a continuous action space. Deep Deterministic Policy Gradient (DDPG) is an actor-critic RL algorithm that extends the capabilities of the DQN to handle continuous action spaces. It uses an actor-critic architecture; and employs a deterministic policy that influences the loss function used for updates. DDPG is an off-policy method where the agent learns from data collected by a separate policy. In this algorithm, the Policy Gradient (PG) part uses gradient descent to optimize parametrized policies concerning the expected return (long-term cumulative reward). The authors in [32] use DDPG for efficient Virtualized Network Functions (VNFs) deployment. Proximal Policy Optimization (PPO) is another type of actor-critic RL algorithm that updates its decision-making policy by collecting a small batch of experiences interacting with the environment. It is an on-policy learning method where the acquired experience samples are only used once to update the current policy. PPO ensures the new policy updates do not deviate too far from the prior ones. The methodology presented in [33] uses information about electricity pricing and renewable energy generation and develops a PPO-based workload management technique to minimize cloud operating costs. The DRL strategies proposed in [32] and [33] minimize the cloud operating costs in a geo-distributed environment but do not consider carbon emissions. However, DRL techniques have disadvantages, including high data and computational intensity, sample inefficiency, exploration and exploitation dilemma, struggle with dynamic environments, and difficulty defining rewards [20]. Therefore, to address the intrinsic shortcomings of using DRL alone, we combine DRL with a Nash equilibrium-based fast non-cooperative game-theoretic technique [17] to propose a novel non-cooperative game-theoretic DRL (GT-DRL). The proposed approach can result in a more adaptive, scalable, and efficient methodology for workload management in geo-distributed data centers.

The application of RL in a multi-agent system is known as Multi-Agent Reinforcement Learning (MARL). Multiple agents typically make judgments based on previous experiences. In this case, an agent, in particular, must understand how to interact with other agents. In [34], the authors investigated methods for connecting various renewable energy generators to geo-distributed data centers from various cloud providers in order to reduce carbon emissions, monetary costs, and service level objectives (SLO) violations caused by renewable energy shortages. The study proposed a MARL-based technique for each data center to estimate the amount of renewable energy to request from each generator. MARL and our proposed GT-DRL framework reflect distinct paradigms for optimizing decision-making processes. MARL uses multiple agents that learn how to interact simultaneously in a shared environment, with a focus on the dynamics that develop from their interactions, which might be cooperative, competitive, or both. In contrast, our GT-DRL is a hybrid technique that optimizes individual agent strategies in a non-cooperative context by combining GT principles with the DRL algorithm. This hybridization in GT-DRL combines DRL's predictive strength with GT's strategic insights to build a meta-heuristic framework that effectively reduces decision-making complexity by simplifying the state-action space. This method improves scalability and efficiency, especially in complicated situations like geo-distributed workload management, by focusing on improving the outcomes for individual agents inside the larger system.

In this work, we study the effects of our proposed GT-DRL on cloud carbon emissions and operating costs. Additionally, we compare our proposed strategy against state-of-the-art DRL methods, including [32] and [33]. Our previous work [17] uses a Nash Equilibrium-based game-theoretic technique for geo-distributed workload management. However, the Nash equilibrium-based game-theoretic technique does not consistently achieve global optima solutions. The scope of this work goes beyond [17] by developing a new carbon emission model that considers the effects of renewable energy generation at different data center locations while making workload distribution decisions. Moreover, we propose a new game-theoretic DRL-based workload management technique that takes a holistic approach to the cloud carbon emissions and operating costs minimization problem, which is shown to be more effective than the multiple resource management strategies presented in [17], [32], and [33]. More details about the comparison strategies are provided in Section 6, and a thorough analysis to evaluate the improvement is presented in Section 7.

## 3 SYSTEM MODEL
### 3.1 Overview

Our proposed framework comprises a geo-distributed-level Cloud Workload Manager (CWM) that distributes incoming workload requests to geographically distributed data centers (Fig. 1). Each data center has its own local Data center Workload Manager (DWM) that takes the workload assigned to it by the CWM and maps requests to compute nodes within the data center. We first describe the system model at the geo-distributed level and then provide further details into the models of components at the data center level.

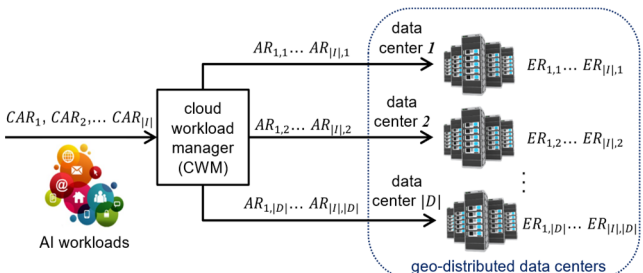

Fig. 1. Cloud workload manager (CWM) performing geo-distributed level task (workload) assignment to data centers.

### 3.2 Cloud Level Model

We consider a rate-based workload management scheme. In our work, an epoch $T^e$ is a discrete one hour time slot. A 24-epoch period represents an entire day. Within the short duration of an epoch, the workload arrival rates can be reasonably approximated as constant. This presumption is based on the fact that the rate of requests or workload coming to a data center frequently appears fairly constant over short time intervals. This is especially true when considering the collective behavior of numerous users or services, where little deviations can average out [35], [36]. CWM makes workload allocation decisions in advance, i.e., before the start of an epoch. As the epoch length is one hour, the workload to resource mapping optimization technique can take up to one hour to calculate the workload distribution strategy.

Let $D$ be a set of $|D|$ data centers, and let $d$ represent an individual data center. We assume that a cloud infrastructure (Fig. 1) comprises $|D|$ data centers, that is, $d = 1,2,...,|D|$ and $d \in D$. Let $I$ be a set of $|I|$ task types, and $i$ represents an individual task type. We consider $|I|$ task (i.e., workload) types, that is, $i = 1,2,...,|I|$ and $i \in I$. A task type $i \in I$ is characterized by its arrival rate and execution rate, i.e., reciprocal of the estimated time required to complete a task of task type $i$ on each of the heterogeneous compute nodes, in each CPU Processor Performance state (P-state). We assume that the beginning of each epoch $\tau$ represents a steady-state scheduling problem. Specifically, at the start of each epoch, the arrival and execution rates for tasks are determined and are assumed to remain constant throughout the epoch. In this problem the CWM splits the cloud level arrival rate $CAR_i(\tau)$ for each task type $i$ into the local data center level arrival rate $AR_{i,d}(\tau)$ and assigns it to each data center $d \in D$. That is,

$$\sum_{d=1}^{|D|} AR_{i,d}(\tau) = CAR_i(\tau), \quad \forall i \in I. \qquad (1)$$

The CWM performs this assignment such that the cloud carbon emissions and operating (energy and network) costs across all data centers are minimized, with the constraint that the execution rates of all task types meet or exceed their arrival rates at each data center, i.e., all tasks complete without being dropped or unexecuted. At the start of each epoch, the CWM calculates (explained in Section 3.3.2) a co-location aware data center maximum execution rate $ER_{i,d}$ for each task type $i$ at each data center $d$. Later, it splits cloud level arrival rate $CAR_i(\tau)$ for each task type $i$ into local data center level arrival rates $AR_{i,d}(\tau)$ such that the data center's maximum execution rate $ER_{i,d}(\tau)$ meets or exceeds the corresponding arrival rate $AR_{i,d}(\tau)$, thus ensuring the workload is completed. That is,

$$ER_{i,d}(\tau) \geq AR_{i,d}(\tau), \quad \forall i \in I, \forall d \in D. \qquad (2)$$

### 3.3 Data Center Level Model
#### 3.3.1 Organization of Each Data Center

Each data center $d$ houses $NN_d$ number of nodes and a cooling system comprised computer room air conditioning (CRAC) units. Let $NCR_d$ be the number of CRAC units. Heterogeneity exists across compute nodes, where nodes vary in their execution speeds, power consumption characteristics, and the number of cores. The number of cores in node $n$ is $NCN_n$.

### 3.3.2 Co-Location-Aware Execution Rates

Tasks competing for shared memory in multicore processors can cause severe performance degradation, especially when competing tasks are memory intensive. The memory intensity of a task refers to the ratio of last-level cache misses to the total number of instructions executed. We use a linear regression model that combines disparate features based on the current tasks assigned to a multicore processor to predict the execution time of a target task $i$ on core $k$ in the presence of performance degradation due to interference from task colocation [37]. These features are, the number of tasks collocated on that multicore processor, the base execution time, the clock frequency, the average memory intensity of all tasks on that multicore processor, and the memory intensity of task $i$ on core $k$ [37].

We classify the task types into memory-intensity classes on each of the node types and calculate the coefficients for each memory-intensity class using the linear regression model to determine a co-located execution rate for task type $i$ on core $k$, $CoER_{i,k}^{core}(\tau)$ [37]. When considering co-location at a data center $d$, the co-location aware data center execution rate for task type $i$ is given by:

$$ER_{i,d}(\tau) = \sum_{n=1}^{NN_d} \sum_{k=1}^{NCN_n} CoER_{i,k}^{core}(\tau). \quad (3)$$

The linear regression model was trained using execution time data collected by executing inference workloads from the AIBench [38] benchmark suite on a set of server-class multicore processors that define the nodes used in our study (see Section 6 for node type details). This model for execution time prediction under co-location interference is derived from real workloads and real server machines and has a Mean Absolute Percentage Error (MAPE) of approximately 7% [37].

### 3.3.3 Data Center Power Model

We use the detailed data center power model from our prior work [16]. Let $CRP_{d,c}(\tau)$ be the CRAC power consumption of CRAC unit $c$ in data center $d$. $NP_n(\tau)$ is node power consumption of node $n$. Each data center is partially powered by either solar power, wind power, or some combination of both. The total renewable power, $RP_d(\tau)$, available at data center $d$ is the sum of the wind and solar power available in epoch $\tau$. $RP_d(\tau)$ can be zero if no renewable power is available. Let $Eff_d$ be an approximation of the power overhead coefficient in data center $d$ due to the inefficiencies of power supply units. $Eff_d$ is always greater than or equal to 1. The total net data center power consumption, $DP_d(\tau)$, at data center $d$, is calculated as:

$$DP_d(\tau) = \left( \sum_{c=1}^{NCR_d} CRP_{d,c}(\tau) + \sum_{n=1}^{NN_d} NP_n(\tau) \right) \cdot Eff_d - RP_d(\tau). \quad (4)$$

For epoch $\tau$, $DP_d(\tau)$ can be negative if the renewable power available at the data center, $RP_d(\tau)$, is greater than the cooling and computing power used in the data center.

### 3.3.4 Data Center Carbon Emission Model

The unique combination of energy sources shapes carbon emissions from electricity consumption in each geographic location. Other factors influencing carbon emissions include the efficiency of power generation and electricity use and the extent of renewable energy generation at a location. As a result, the emissions can fluctuate hourly, daily, monthly, and annually. The U.S. Energy Information Administration (EIA) provides estimates of these emissions on a monthly and annual basis. In 2021, power plants burning coal, natural gas, and petroleum fuels were responsible for around 61% of the total yearly U.S. utility-scale electricity net generation. However, they accounted for 99% of U.S. carbon emissions associated with utility-scale electric power generation. EIA publishes annual carbon emissions and average annual carbon emissions factors related to total electricity generation by the electric power industry in the United States and each state [39]. These carbon factors, $C_d^{factor}$, are in kilograms of carbon emitted per kWh. And the value differs for every data center location. The data center carbon emissions $DE_d(\tau)$ at data center $d$ can be defined as:

$$DE_d(\tau) = C_d^{factor} \cdot DP_d(\tau). \quad (5)$$

### 3.3.5 Net Metering Model

Net metering allows data center operators to sell back the excess renewable power generated on-site to the utility company. When the extra power is added to the grid, utility companies pay a fraction of the retail price. The value of this factor varies by data center location. This fraction is called the net metering factor, $\alpha_d$.

### 3.3.6 Peak Power Demand Model

Most utility providers charge a flat-rate (peak demand) fee based on the highest (peak) power consumed at any instant during a given billing period, e.g., a month. The peak demand price per $kW$ at data center $d$ is denoted as $P_d^{price}$. We define $Pc_d^{peak}(\tau)$ as the highest grid power consumed since the beginning of the current month, including the current epoch $\tau$. We define $Pp_d^{peak}(\tau)$ as the highest grid power consumption since the beginning of the current month until the start of the current epoch $\tau$. The peak power cost increase at data center $d$, $\Delta_d^{peak}(\tau)$, is then defined as:

$$\Delta_d^{peak}(\tau) = P_d^{price} \cdot \left( Pc_d^{peak}(\tau) - Pp_d^{peak}(\tau) \right) \quad (6)$$
if $Pc_d^{peak}(\tau) \geq Pp_d^{peak}(\tau)$, else it is equal to 0.

The peak power cost increase $\Delta_d^{peak}(\tau)$ is calculated in each epoch $\tau$ and summed over all epochs in a billing period to calculate the total peak power cost.

### 3.3.7 Network Cost Model

Tasks and their associated data migrated across data centers incur a network cost. Two principal components of the network cost are the network price per data traffic unit (\$/GB), $N^{price}$, and the amount of data volume (GB) for the number of tasks migrated (outward). We assume that $N^{price}$ is the same across all data centers. Let $S_i$ be the size (in GB)

of a task type $i$. The network cost at data center $d$, $NC_d(\tau)$, is calculated as

$$NC_d(\tau) = \sum_{i=1}^{|I|} \sum_{n=1}^{NN_d} (N^{price} \cdot S_i). \quad (7)$$

### 3.3.8 Data Center Cost Model

The electricity price per $kWh$ at data center $d$ is defined as $E_d^{price}(\tau)$. Data center operators can use net metering if the total net power consumed throughout the data center, $DP_d(\tau)$, is negative. For such conditions, the total data center cost $DC_d(\tau)$ for data center, $d$ can be defined as

$$DC_d(\tau) = E_d^{price}(\tau) \cdot \alpha_d \cdot DP_d(\tau) + \Delta_d^{peak}(\tau) + NC_d(\tau) \quad (8)$$

where $\alpha_d = 1$ if $DP_d(\tau)$ is positive and $0 \leq \alpha_d \leq 1$ otherwise. The first term in (8) represents the TOU electricity cost, the second term represents the peak demand cost, and the third represents the network cost.

## 4 PROBLEM FORMULATION

### 4.1 Overview

We consider a scenario with multiple data centers maintained by a cloud service provider. The system is assumed to be under-subscribed in the sense that the system is expected to have enough computation resources to complete the workload without requiring that any tasks be dropped or terminated before completion i.e., satisfy constraint defined by (2). We consider AI inference batch workload where the tasks originate off-site from the data centers, and we do not consider the transmission time (latency) and cost from a task origin to a data center. If the CWM migrates a task from one location to another, a data transfer cost is associated with it (as discussed in Section 3.3.7). The main objective of a CWM is to allocate the workload across geo-distributed data centers to minimize cloud carbon emissions (the sum of (5) across all data centers). We also consider the related problem of minimizing the monetary cloud operating (energy and network) costs of the system (the sum of (8) across all data centers). Our proposed technique ensures that all workloads are completed according to the constraint defined by (2).

Allocating heterogeneous tasks to data centers with varying aisle, CRAC, and node configurations, heterogeneous servers, and CPU core counts, each with a unique P-state, adds significant complexity. When you include co-location execution, dynamic electricity pricing, inter-data center data transfer costs, peak demand prices, net metering factors, and carbon factors per location, along with execution rate constraints, the problem becomes more complex. These factors collectively contribute to the NP-hardness of the problem, as they exponentially increase the number of potential solutions, making a simple, linear approach or exhaustive search impractical. To tackle this, we propose a game-theoretic DRL-based workload management framework. We model the geo-distributed level workload distribution problem as a non-cooperative game. In a non-cooperative game, there could be a finite (or infinite) number of players who aim to maximize/minimize their objective independently but ultimately reach an equilibrium. For a limited number of players, this equilibrium is called the Nash equilibrium [17].

### 4.2 Objective Functions

The main goal of our proposed framework is to minimize aggregate cloud data center carbon emissions. Alternatively, our framework allows updating the objective function to optimize cloud operating (energy and network) costs. This work considers single objective optimization and does not optimize carbon emissions and operating costs together.

Let $J_d$ be a set of node types in a data center $d$, and let $j$ represent an individual node type. In a data center $d$, there are a total of $NN_{d,j}$ nodes of node type $j$. Let $P_j^D$ be the average peak dynamic power for node type $j$. It is calculated by averaging (for all task types) the peak power for each task type $i$ executing on node type $j$. Let $DP_d^{max}$ be the maximum data center power consumption possible at data center $d$, which can be calculated as:

$$DP_d^{max} = \left( NCR_d \cdot CRP_{d,c}^{max} + \sum_{j \in J_d} NN_{d,j} \cdot P_j^D \right) \cdot Eff_d \quad (9)$$
$$- RP_d(\tau).$$

Recall that $RP_d(\tau)$ is the total renewable power available at the data center $d$. Let $DP_{i,d}^{est}(AR, \tau)$ be the estimated data center power consumption possible for each task type $i$ at data center $d$, which can be calculated as:

$$DP_{i,d}^{est}(AR, \tau) = \left( \frac{DP_d^{max} \cdot AR_{i,d}(\tau)}{ER_{i,d}(\tau)} \right). \quad (10)$$

Observe that, at data center $d$ with zero $RP_d(\tau)$ for each task type $i$, $DP_{i,d}^{est}(AR, \tau)$ in (10) will increase to $DP_d^{max}$ as the data center level arrival rate $AR_{i,d}(\tau)$ approaches to its maximum execution rate $ER_{i,d}(\tau)$. Thus, we can say that $DP_{i,d}^{est}(AR, \tau)$ is function of $AR$ and $\tau$.

### 4.2.1 Estimated Cloud Carbon Emissions

The estimated data center carbon emissions $DE_{i,d}^{est}(AR, \tau)$ incurred by the task type $i$ at data center $d$ with that data center maximum execution rate $ER_{i,d}(\tau)$, can be calculated by modifying (5) as:

$$DE_{i,d}^{est}(AR, \tau) = C_d^{factor} \cdot DP_{i,d}^{est}(AR, \tau). \quad (11)$$

The estimated cloud carbon emissions incurred for the task type $i$, $CET_i^{est}(AR, \tau)$, can be calculated as:

$$CET_i^{est}(AR, \tau) = \sum_{d=1}^{|D|} DE_{i,d}^{est}(AR, \tau) \quad (12)$$

Our optimization goal is to minimize the estimated cloud carbon emissions incurred for all task types, $CE^{est}(AR, \tau)$, which can be calculated as:

$$CE^{est}(AR, \tau) = \sum_{i=1}^{|I|} CET_i^{est}(AR, \tau) \quad (13)$$

subject to the constraints defined by (1) and (2).

### 4.2.2 Estimated Cloud Operating Cost

Let $NC_{i,d}^{max}$ be the maximum network cost possible for each task type $i$ at data center $d$, which can be calculated as:

$$NC_{i,d}^{max} = N^{price} \cdot NN_d \cdot S_i. \quad (14)$$

Let $NC_{i,d}^E(AR, \tau)$ be the estimated network cost possible for each task type $i$ at data center $d$, which can be calculated as:

$$NC_{i,d}^{est}(AR, \tau) = \frac{NC_{i,d}^{max} \cdot AR_{i,d}(\tau)}{ER_{i,d}(\tau)}. \qquad (15)$$

Observe that, at data center $d$ for each task type $i$, $NC_{i,d}^{est}(AR, \tau)$ in (15) will increase to $NC_{i,d}^{max}$ as the data center level arrival rate $AR_{i,d}(\tau)$ approaches to its maximum execution rate $ER_{i,d}(\tau)$. Thus, we can say that $NC_{i,d}^{est}(AR, \tau)$ is a function of $AR$ and $\tau$.

Then the estimated data center cost $DC_{i,d}^{est}(AR, \tau)$ incurred by the task type $i$ with a data center maximum execution rate $ER_{i,d}(\tau)$ at data center $d$, can be calculated by modifying (8) as:

$$DC_{i,d}^{est}(AR, \tau) = E_d^{price}(\tau) \cdot \alpha_d \cdot DP_{i,d}^{est}(AR, \tau) + \Delta_d^{peak}(\tau) + NC_{i,d}^{est}(AR, \tau) \qquad (16)$$

where $\alpha_d = 1$ if $PD_{i,d}^{est}(AR, \tau)$ is positive and $0 \leq \alpha_d \leq 1$ otherwise.

The estimated cloud operating costs incurred for the task type $i$, $CCT_i^{est}(AR, \tau)$, can be calculated as:

$$CCT_i^{est}(AR, \tau) = \sum_{d=1}^{|D|} DC_{i,d}^{est}(AR, \tau) \qquad (17)$$

Our (second) optimization goal is to minimize the estimated cloud operating costs incurred for all task types, $CC^{est}(AR, \tau)$, can be calculated as:

$$CC^{est}(AR, \tau) = \sum_{i=1}^{|I|} CCT_i^{est}(AR, \tau) \qquad (18)$$

subject to the constraints defined by (1) and (2).

## 5 OUR APPROACH

In this section, we formulate the cloud workload distribution problem as the Nash equilibrium-based non-cooperative game. Then, we combine it with DRL to design a new intelligent game-theoretic DRL (GT-DRL). The choice of non-cooperative game theory over traditional optimization is motivated by the requirement to model competitive environments in which several independent agents pursue their own goals. Similarly, while DRL has shown useful in learning optimal policies for individual agents, it may not completely account for the strategic interactions of several agents, particularly in competitive or non-cooperative environments. Our GT-DRL approach produces more realistic rapid decision-making models, especially in cases where collaboration is not possible and agents' actions are interdependent. In our workload management problem, the goal is to split the cloud level arrival rate $CAR_i(\tau)$ for each task type $i$ into local data center level arrival rates $AR_{i,d}(\tau)$ and assign them to each data center $d \in D$. The following subsections describe the components of our technique and its architecture in more detail.

### 5.1 Game Theoretic Component

We model our cloud workload management problem as a non-cooperative game played among a set of players. Fig. 2 shows the architecture of the game-theoretic component, where the dotted yellow boxes show the elements of our non-cooperative game. These elements are:

- **Players:** a finite set of players, where each task type $i$ is a player; $i = 1, 2, \ldots, |I|$ (as per Section 3.2)
- **Strategy sets:** load distribution strategy of a player $AR_i = \{AR_{i,1}, AR_{i,2}, \ldots, AR_{i,|D|}\}$
- **Rewards:** the goal of each player $i$ is to minimize rewards associated with its strategy, i.e., the cloud carbon emissions incurred $CET_i^{est}(AR, \tau)$ calculated using (12), or cloud operating costs incurred $CCT_i^{est}(AR, \tau)$ calculated using (17) associated with its strategy, subject to the constraints defined by (1) and (2).

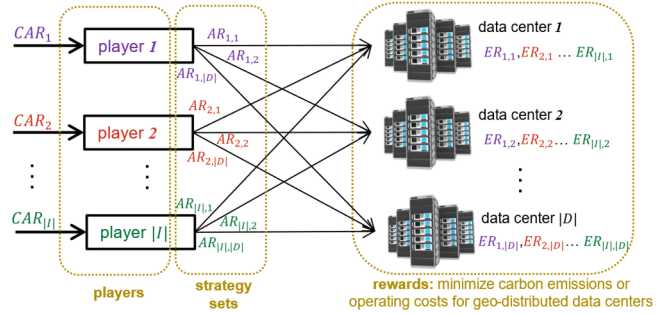

Fig. 2. Game-theoretic component of our framework

The game described here can be solved using a Nash equilibrium to obtain the workload distribution strategy for the cloud data centers. For minimizing carbon emissions, the Nash equilibrium of our non-cooperative game is a load distribution strategy of the entire cloud $AR = \{AR_1, AR_2, \ldots, AR_{|I|}\}$ such that for each player $i$:

$$AR_i \in \underset{AR_i}{argmin}\, CET_i^{est}(AR, \tau). \qquad (19)$$

subject to the constraints defined by (1) and (2).

For minimizing cloud operating costs, the Nash equilibrium of our non-cooperative game is a load distribution strategy of the entire cloud $AR = \{AR_1, AR_2, \ldots, AR_{|I|}\}$ such that for each player $i$:

$$AR_i \in \underset{AR_i}{argmin}\, CCT_i^{est}(AR, \tau). \qquad (20)$$

subject to the constraints defined by (1) and (2).

If no player can further minimize the reward (emissions or costs) incurred by changing its current strategy to another one, the strategy $AR$ is a Nash equilibrium. In this equilibrium, a player $i$ cannot decrease the overall reward incurred by choosing a different workload distribution strategy $AR_i$ given the other players' workload distribution strategies. Note that the goal of our approach is to find a workload distribution strategy for the entire cloud $AR = \{AR_1, AR_2, \ldots, AR_{|I|}\}$ where each player $i$ determines its workload distribution strategy $AR_i = \{AR_{i,1}, AR_{i,2}, \ldots, AR_{i,|D|}\}$ that satisfies equations (19) or (20). To find these strategies, our previous implementation of the Nash equilibrium-based game-theoretic technique used the "Best-Reply" algorithm [17]. This mathematical algorithm that often fails to achieve global optimum solutions. To overcome this issue and improve the quality of our optimization approach, we use DRL for each player $i$ to determine its workload distribution strategy $AR_i$.

## 5.2 DRL Component

DRL uses DNNs with several layers to explore a problem space. DNNs are highly suited for analyzing complex problems and high-dimensional information because they can autonomously learn hierarchical features. Our cloud workload scheduling problem for heterogeneous geo-distributed data centers is complex. Therefore, we consider a DRL-based method to automatically discover an optimal workload distribution strategy. Since DRL agents can sample from an environment or simulator, they need a reward function as input. As a result, they can interact with the environment, receive rewards or penalties for their actions, and learn optimal policy through trial and error.

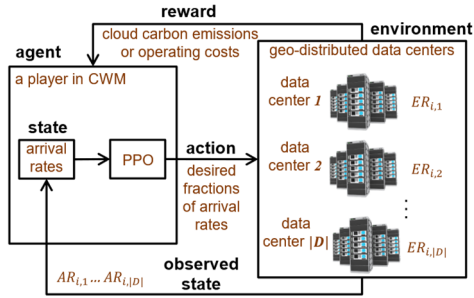

Fig. 3. DRL component of our framework

Fig. 3 displays the structure of the DRL component of our framework. The game-theoretic component of our framework uses DRL to find workload distribution strategy $AR_i = \{AR_{i,1}, AR_{i,2}, \ldots, AR_{i,|D|}\}$ for each player $i$ (as explained in Section 5.1). In this work, we refer to the state as a strategy, in alignment with the strategy set defined in the previous section. The strategy or state, represented by a set of arrival rates, forms the basis of the current state. Agents (or players as defined in GT component) take actions, which consist of desired fractions, that predict the future or next strategy. By distributing the workload, characterized by their arrival rates, these actions guide the system towards an optimal state or strategy. Predictions made by different agents/players are independent on each other because of the non-cooperative nature of the system. Below are the major elements of the DRL component.

- **Agent:** decision maker and learner, i.e., a player or a task type $i$ in CWM.
- **Environment:** information about geo-distributed data centers such as cloud level arrival rates for each task type, data center maximum execution rates for each task type at all data centers, maximum data center power consumptions, renewable power availability, carbon factors, electricity prices, net-metering factors, peak prices for all data center locations, and network costs for each task type at all data centers.
- **State:** state of the agent in the environment, i.e., current arrival rate distribution of a player/task type (i.e., the strategy set as described in Section 5.1)
- **Reward:** feedback value from the environment to assess the quality of the prior action, i.e., optimization objective of a player (i.e., reward as described in Section 5.1). The reward function calculates these values using information available in the environment. Typically, an agent tries to maximize the rewards; but our optimization goal is to minimize the rewards. Thus, we consider negative values.
- **Actions:** a set of $|D|$ desired fraction values. Each data center $d$ gets a desired fraction $DF_{i,d}(\tau)$ of cloud level arrival rate $CAR_i(\tau)$ for each task type $i$. This condition is maintained at the beginning of every iteration. The algorithm chooses a set of actions that satisfy this condition. Thus, the local data center level arrival rate $AR_{i,d}(\tau)$ for a task type $i$ at a data center $d$, can be calculated as,

$$AR_{i,d}(\tau) = DF_{i,d}(\tau) \cdot CAR_i(\tau)$$
$$\text{where} \sum_{d=1}^{|D|} DF_{i,d} = 1, \quad \forall i \in I \quad (21)$$

subject to the constraints defined by (1) and (2).
- **Policy:** The agent's policy for determining the next course of action based on the current state.
- **Value function:** The predicted long-term reward with discount, in contrast to the short-term reward.
- **Action-value:** The long-term reward of the current state, taking action under a policy.

## 5.3 GT-DRL Framework

We adapt the Proximal Policy Optimization (PPO) DRL algorithm in our framework. It is an on-policy method designed to address the challenges of other policy optimization methods, such as the difficulty of choosing a good step size and poor sample efficiency. It is an actor-critic method where the critic's learning is used to benefit the actor. A value function that the critic learns is used to calculate the expected return (sum of future rewards) from a given state. The actor uses these estimates to modify its policy. In our proposed GT-DRL, a PPO-based DRL is combined with non-cooperative game theory to form a game-theoretic DRL, as shown in Fig. 4.

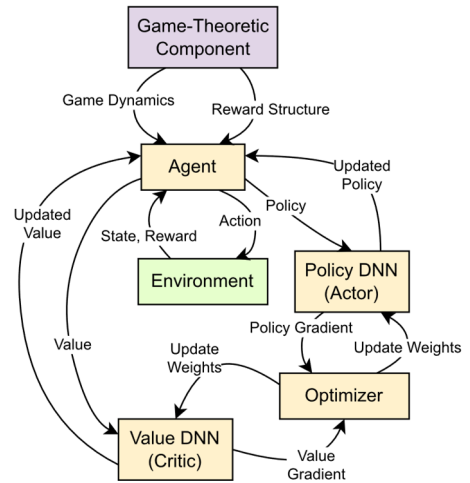

Fig. 4. GT-DRL Framework

Fig. 4 shows the flow of our GT-DRL framework. Different colors represent different components. All yellow boxes are part of the PPO algorithm used by the agent. The agent interacts with the environment and receives the state and reward. It is an iterative optimization process that allows an agent to learn and improve its strategy for decision-making in an environment. It uses the policy DNN (actor) to decide what action to take in the environment. It also utilizes the value DNN (critic) to estimate the value function of the current state. The policy gradient from the actor and the value gradient from the critic are evaluated and passed to the optimizer. The PPO algorithm refines and updates both the actor and the critic DNNs using the optimizer. The policy and the value function are improved based on the observed rewards and the next state from the environment. The agent then employs the updated/improved policy for the next interaction with the environment. The game-theoretic component (as described in Section 5.1) is used to apply game dynamics and to provide the reward structure to the agent so that the agent determines the optimal workload distribution strategy. The game theory guides the exploration and improves the effectiveness of the policy. Breaking the complex problem of geo-distributed workload management into smaller chunks makes PPO more efficient.

The DRL component of our GT-DRL finds the workload distribution strategy $AR_i = \{AR_{i,1}, AR_{i,2}, ..., AR_{i,|D|}\}$ for each player $i$ first and then combines them to form a workload distribution strategy of the entire cloud $AR = \{AR_1, AR_2, ..., AR_{|I|}\}$. On the other hand, if deployed directly in our geo-distributed environment, a conventional (non-game-theoretic) DRL would find the optimal workload distribution strategy of the entire cloud $AR = \{AR_{1,1}, ..., AR_{i,d}, ..., AR_{|i|,|D|}\}$ such that the cloud carbon emissions $CE^{est}(AR, \tau)$ calculated using (13) or cloud operating costs $CC^{est}(AR, \tau)$ calculated using (18) are minimized. In the conventional DRL, the set of states, i.e., the distribution strategy $AR$ involved in the optimization, has $|I| \times |D|$ variables. In our GT-DRL each player $i$ minimizes the reward for its own strategy. Therefore, the set of states, i.e., the distribution strategy $AR_i$ involved in this optimization, has only $|D|$ variables. The non-cooperative game-theoretic nature of GT-DRL reduces the state and action sets from $|I| \times |D|$ variables to $|D|$ variables. This greatly reduces the state-action space and makes the underlying DNN fast and lightweight.

## 6 SIMULATION ENVIRONMENT

We compare our proposed GT-DRL approach to mathematical (e.g., greedy, genetic), game-theoretic, and other DRL-based techniques. We implemented the DRL-based techniques using Python 3.8, OpenAI Gym [40], and PyTorch [41]. In addition, we used Stable Baseline 3 (SB3) [42], a set of reliable implementations of RL algorithms in PyTorch. It is a widely used, open-source, and well-tested RL library. We run the algorithm training in an offline environment because training in a real setup is too slow and often infeasible for large complex environments. This is a common practice in DRL-based optimization research [43], [33]. We train our DRL models with randomly sampled values of uniformly distributed set of arrival rates. All other environment variable values that we use are detailed in the subsequent portion of this section. We monitor the rewards and train the DRL models until either the rewards converge, or no further improvements are observed. The comparison techniques for geo-distributed data center management considered in our experimental analysis are summarized below.

**(a) Force-Directed (FD):** This is a greedy approach proposed in [18]. It is a variation of force-directed scheduling, frequently used for optimizing semiconductor logic synthesis.

We adapt this approach to our environment and implement FD. It is an iterative method that selectively performs operations to minimize system forces until all constraints are met.

**(b) Genetic Algorithm (GA):** We adapt the GA from [16]. The Genitor style [44] GA has two parts: a genetic algorithm-based CWM and a local data center level greedy heuristic that is used to calculate the fitness value of the genetic algorithm. The local greedy heuristic has information about task-node power dynamic voltage and frequency scaling (DVFS) models described in [16].

**(c) Nash equilibrium (NASH):** This technique, which was proposed in [17], integrates detailed models for data center power consumption, cost, and inter-data center network. This paper uses Nash equilibrium-based game-theoretic workload distribution techniques for geo-distributed data centers.

All the above methods (a), (b), and (c) are mathematical and need to be used with some workload prediction systems to avoid delays with workload allocation. This limitation does not apply to the DRL-based techniques because of their dynamic nature.

**(d) Deep Deterministic Policy Gradient (DDPG):** We do not compare our proposed technique with DQN-based prior works [26], [27], [28], [29], [30], [31] because DQN-based algorithms use discrete action space while our complex optimization problem uses continuous action space. Instead, we compare it against DDPG [32], which extends DQN to the continuous action space. We adapt the framework developed in [32] to our environment and build a DDPG-based DRL.

**(e) Proximal Policy Optimization (PPO):** PPO adjusts its policies less frequently and more cautiously. It adds a "clip" function to the policy objective function to avoid the policy altering too much with one update. We modify the PPO-based DRL proposed in [33] to our problem and environment.

Experiments were conducted for three geo-distributed data center configurations containing four, eight, and sixteen data centers. As shown in Fig. 5, the locations of the data centers in the three configurations were selected from major cities around the continental United States to provide a variety of wind and solar conditions among sites and at various times of the day. The locations of each configuration were selected so that each configuration would have an even east-coast to west-coast distribution to exploit better TOU pricing, peak demand pricing, net metering, renewable power, and carbon factor. Each data center comprises 4,320 nodes arranged in four aisles and is heterogeneous, with nodes from two or three node types, with most locations having three node types. We use Intel Xeon E3-1225v3 (4 core), Xeon E5649 (6

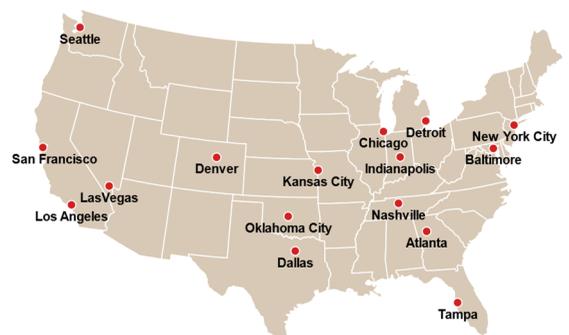

Fig. 5. Location of simulated data centers

core), and Xeon E5-2697v2 (12 core) processor-based nodes.

The time of each epoch $\tau$ was set to be one hour. We limit our mathematical techniques (FD, GA, and NASH) to run for a maximum of one hour (epoch length). The DRL-based techniques run in a few seconds and can be deployed at run-time because of their dynamic nature. Carbon factor values used in the experiments are taken from EIA [39]. The network prices used in the experiments were taken from Amazon Web Services [45]. The real-world TOU electricity prices and the peak demand prices were taken directly from the utility companies. We calculated the renewable power at each data center location using solar and wind data from the National Solar Radiation Database [46]. For all experiments, we used the renewable power calculated for the month of June. Fig. 6 depicts the renewable power available at four locations with the yellow bar plots. The left y-axis is normalized to the highest renewable power available. The net metering factor $\alpha_d$ is 1 at most locations; it is less than 1 in rare cases; and it is 0 in some cases, i.e., net metering is not available at that location [47].

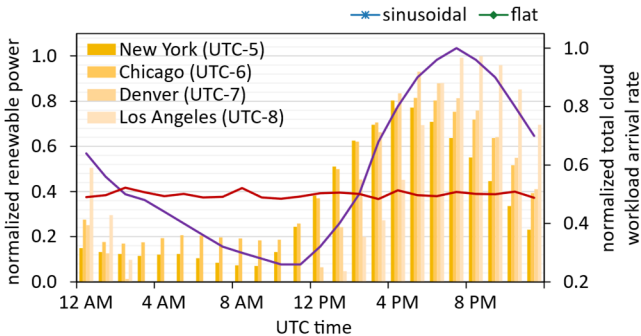

Fig. 6. Renewable power and workload arrival rate at four data centers

Organizations are heavily using AI inference workloads as discussed in Section 1. Due to the popularity of such workloads among cloud providers, we use data-intensive AI inference workloads from the AIBench [38] benchmark suite. Table 2 summarizes the task types, ten in total, from this suite that we consider in our work. Task execution times and co-located performance data for tasks of the different memory intensity classes were obtained from running the benchmark applications on the three heterogeneous nodes. We considered two different workload arrival rate patterns: sinusoidal and flat. Fig. 6 depicts a sinusoidal task arrival rate pattern and a flat task arrival rate pattern with line plots. The right y-axis is normalized to the highest total cloud workload arrival rate. A sinusoidal task arrival rate pattern exists in environments where the workload traffic depends on consumer interaction and follows their demand during the day. However, for the environments where continuous computation is needed, and the workload pattern is non-user/consumer interaction specific, the task arrival rate pattern is usually flat (nearly constant). We conducted five simulation runs for every experiment with different workload arrival patterns. For each run, the arrival rate values were randomly sampled (with a normal distribution). We used the original arrival rate values, shown in Fig. 6, as the mean and employed 20% of the mean as the standard deviation.

## 7 EXPERIMENTS

### 7.1 Cloud Carbon Emissions Comparison

Our first set of experiments analyzes the cloud carbon emissions, the total amount of carbon emissions across all data centers in the cloud service provider's infrastructure, for each technique described in Section 6. For this experiment, the optimization objective of all methods is to minimize cloud carbon emissions. This experiment used a data center configuration with four locations running a sinusoidal workload pattern. We conducted five simulation runs and analyzed each technique's cloud carbon emissions variation over a day by plotting the average with the standard error bars, as shown in Fig. 7. The y-axis is normalized to the highest cloud operating cost.

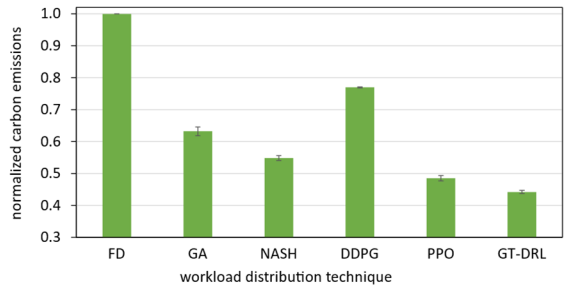

Fig. 7. Cloud carbon emissions for each technique over a day for a configuration with four data centers

From Fig. 7, it can be observed that the FD method performed the worst, severely over-provisioning nodes. Over-provisioning caused nodes to consume high power, increasing data center carbon emissions. The GA method has information about task-node power (DVFS) models [16], allowing it to make better task placement decisions. NASH performed the best among mathematical methods (FD, GA, and NASH). Because of the non-cooperative nature of this method [17], each player/task type determined the lowest possible carbon emissions with its workload allocation strategy.

It can be observed that the DDPG technique did not perform well among DRL methods (DDPG, PPO, and GT-DRL). DDPG depends on exploration to find the optimal policies. Our complex objective function optimization frequently demands a thorough search through the solution space, which is unsuitable for DDPG's exploration strategy. On the other hand, PPO urges the acting policy to adhere to the prior policy as closely as possible while permitting exploration within a specific range. Due to its improved sample efficiency, PPO is better suited to optimization problems like ours, where costly exploration is involved. Finally, GT-DRL performs the best because it allows multiple players to quickly optimize policy in a non-cooperative way, covering a broader solution space.

TABLE 2: Task Types Used in Experiments

| Benchmark | Algorithm | Dataset |
|---|---|---|
| Image Classification | ResNet50 | ImageNet |
| Image Generation | WassersteinGAN | LSUN |
| Image-to-Text | Neural Image Caption Model | Microsoft COCO |
| Image-to-Image | CycleGAN | Cityscapes |
| Speech Recognition | DeepSpeech2 | Librispeech |
| Face Embedding | Facenet | VGGFace2, LFW |
| 3D Face Recognition | 3D Face Model | Intellifusion |
| Video Prediction | Motion-Focused Predictive Model | Robot Pushing |
| Image Compression | Recurrent Neural Network | ImageNet |
| 3D Object Reconstruction | Convolutional Encoder-Decoder Network | ShapeNetCore |

## 7.2 Data Center Scalability Analysis

In this experiment, we analyze the impact of larger problem sizes. Five simulation runs were conducted with the sinusoidal workload patterns for 8 and 16 data center configurations and the previously discussed 4 data center configuration. Fig. 8 shows the % cloud carbon emissions reduction of GT-DRL over each technique from each configuration, where the y-axis is the % cloud carbon emissions reduction.

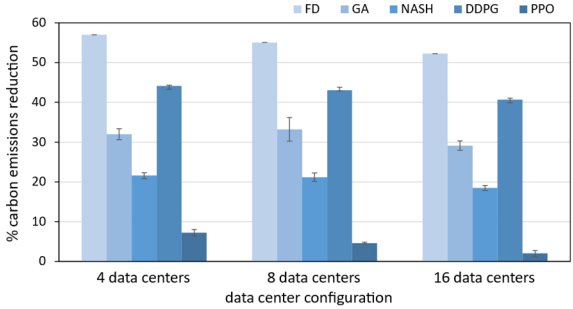

Fig. 8 % cloud carbon emissions reduction of GT-DRL over each technique for a configuration with 4, 8, and 16 data centers

As the number of data centers in the group grows, the problem size increases and becomes more complex. Therefore, the results show that cloud carbon emissions reduction decrease with the increasing number of data centers. The large error bars for the GA indicate a higher degree of variability/sensitivity where GA did not produce consistent results and was unstable within the given time constraint (one hour). For DRL-based methods, the state and action space increase for large problems, increasing the problem complexity and making the DRL-based methods difficult to converge. For our GT-DRL, the rise in problem space makes it difficult for players to find equilibrium points or optimal strategies that balance the actions of different players to achieve the common objective, i.e., to minimize the overall cloud carbon emissions. These experiments confirm that our technique, GT-DRL, consistently performed the best for all problem sizes. The results from Fig. 8 show that, on average, GT-DRL achieved up to 55%, 32%, 21%, 43%, and 5% better cloud carbon emissions savings than FD [18], GA [16], NASH [17], DDPG [32], and PPO [33] respectively.

## 7.3 Workload Arrival Rate Pattern Analysis

For this experiment, we considered a configuration with eight data centers executing both sinusoidal and flat workload arrival patterns. Fig. 9(a) and Fig. 9(b) show both arrival rate patterns (shaded yellow color). We conducted five simulation runs for each workload pattern. The experiment analyzes the cloud carbon emissions at each epoch (one-hour interval) over a day. The left y-axis is normalized to the highest carbon emitted at any epoch, and the right y-axis is normalized to the highest total cloud workload arrival rate. Recall that each task type is characterized by its arrival rate and the estimated time required to complete the task on each heterogeneous compute node. The workload management techniques distribute the workload to minimize cloud carbon emissions across all data centers with the constraint that the execution rates of all task types meet their arrival rates (2). Therefore, the workload assignment was altered by changing the incoming arrival rate pattern, which further affected cloud carbon emissions. The results shown in Fig. 9(a) and Fig. 9(b) indicate that the geo-distributed system responded differently to sinusoidal and flat arrival rate patterns.

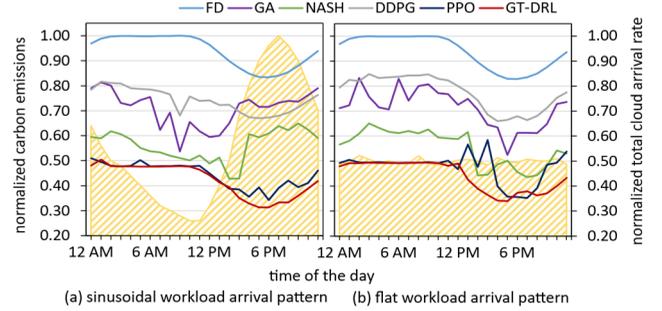

Fig. 9. Comparison of cloud carbon emissions among techniques over a day for (a) sinusoidal and (b) flat workload arrival rate patterns for eight data centers

Overall, cloud carbon emissions were higher for the sinusoidal workload arrival pattern, where the elevated workload arrival rates in the second half of the day produced higher resource demand. This high resource demand resulted in higher power consumption causing more carbon emissions than the flat workload arrival pattern. For sinusoidal workload arrival in Fig. 9(a), the PPO-based DRL techniques performed better than all others during the high workload traffic period. The GT-DRL consistently performed the best for both arrival rate patterns during the day. This shows that our proposed technique allocated resources most effectively to reduce power consumption that reduced carbon emissions. For both arrival rate patterns, we see a dip in carbon emissions around 6 PM UTC because of the increased availability of renewable power (as shown in Fig. 6). We analyze the effects of renewable power in the next section.

## 7.4 Renewable Power Sensitivity Analysis

We performed another experiment to analyze the impact of the amount of renewable power on cloud carbon emissions. As mentioned in Section 3.3.4, carbon emissions from electricity consumption at each data center depend on renewable power generation. As a result, the emissions

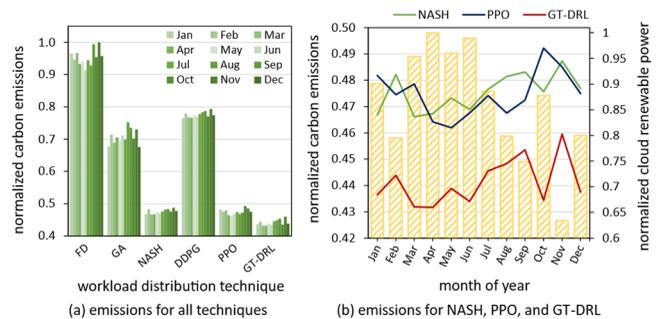

Fig. 10. Impact of renewable power on cloud carbon emissions among techniques over a year for (a) all techniques and (b) NASH, PPO, and GT-DRL for eight data centers

can fluctuate hourly, daily, monthly, and annually. This experiment considered all techniques for eight data centers executing a sinusoidal workload. We simulated 12 runs, where each run used different renewable power available at each month of the year. Results from Fig. 10(a) show cloud carbon emissions for all techniques over each month of the year. The y-axis is normalized to the highest carbon emitted by any technique at any month of the year. Each column shows the amount of carbon emitted for a different month. Here, each technique followed the performance trend we discussed in our first experiment, where FD performed the worst, followed by DDPG and GA. GT-DRL performed the best, followed by PPO and NASH. However, NASH, PPO, and GT-DRL came close to each other. To closely analyze the performance difference and impact of those techniques for each month, we plotted them differently on a finer scale, as shown in Fig. 10(b). The left y-axis is normalized to the highest carbon emitted at any month, and the right y-axis is normalized to the highest total renewable power available (shaded yellow bars) at all data centers combined. The results from Fig. 10(b) show that the cloud carbon emissions decreased with the increase in renewable power available. The plot showed that our proposed GT-DRL utilized the available renewable power most effectively and consistently performed the best for all months of the year.

## 7.5 Cloud Operating Costs Comparison

For each technique, this experiment analyzes the cloud operating costs, comprised of energy and network costs associated with inter-data center data transfers. The optimization objective for all methods in this experiment is to minimize cloud operating costs. This experiment used a data center configuration with four locations running a sinusoidal workload pattern. We conducted five simulation runs with different sinusoidal workload arrival patterns and analyzed the cloud operating costs that vary for each technique over a day. Results are shown in Fig. 11(a), where each stacked column/block represents the cloud operating costs for an epoch. The column values are normalized to the highest cost value at any epoch in the day. The results show that the costs for each method are very high during the first epoch (12 AM – 1 PM) because the period for which the results are shown represents the first day of the month where the initial peak demand cost is added. This effect stays there for the first day and would not be present for other days of the month. Although the first epoch costs are significant for FD, the DDPG performed the worst. Because of its off-policy nature, DDPG learns from different experiences than those it uses to make decisions. Such highly unstable learning and non-reassuring convergence made it challenging to obtain good solutions, resulting in high costs for most of the epochs. GA performed the worst among mathematical methods (FD, GA, and NASH). Similar to DDPG, the GA took a long time to explore the solution space and could not perform an adequate number of GA generations that could take place within the time limit (one hour by default). FD performed better than GA and DDPG. Because of its greedy nature, it completed many more iterations to explore the solution space. But it could not perform better than NASH and PPO-based DRL methods because it tends to get stuck in local minima [17]. Using its game-theoretic decision-making, NASH produced better results than DDPG, GA, and FD. PPO is an on-policy algorithm that prevents significant policy updates, ensuring stability and continuous improvement in the objective function. Having a better convergence makes PPO suitable for our geo-distributed workload optimization problem. Finally, our proposed technique, i.e., GT-DRL, performed better than other algorithms. In our proposed game-theoretic approach, the Nash equilibrium concept brought stability to the system by reaching states where no single player could unilaterally improve its performance (after reaching equilibrium). This resulted in a more robust solution in fluctuating geo-distributed workload scheduling environments.

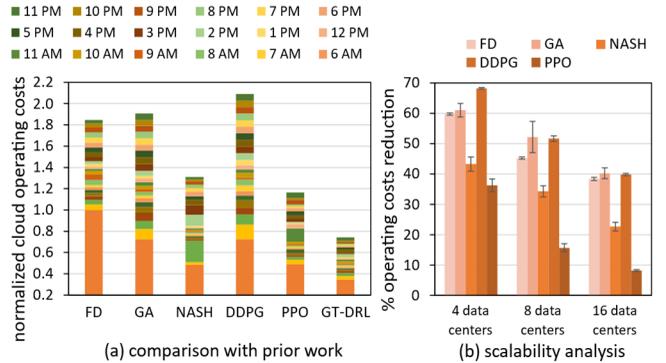

Fig. 11. (a) Epoch-based cloud operating costs for each technique over a day for a configuration with four data centers and (b) % cloud operating costs reduction of GT-DRL over each technique for a configuration with 4, 8, and 16 data centers

Fig. 11(b) shows the scalability analysis regarding the % cloud operating costs reduction of GT-DRL over each technique from each configuration, where the y-axis is the % cloud operating costs reduction. The results show that cloud operating cost reduction decrease with the increasing number of data centers because of the increasing complexity of the problem. The GA performed the worst for 8 and 16 data centers. As the problem complexity increases, the number of GA generations that can take place within the time limit (one hour by default) decreases, which decreases the performance of GA. GT-DRL performed the best overall. The results from Fig. 11(b) show that, on average, GT-DRL achieved up to 48%, 51%, 34%, 53%, and 20% better cloud operating costs savings than FD [18], GA [16], NASH [17], DDPG [32], and PPO [33] respectively.

## 8 LIMITATIONS AND FUTURE DIRECTIONS

This study primarily focuses on the mathematical and simulation-based analysis of our proposed cloud workload distribution technique. Future research can focus on deploying and testing our proposed approach in operational cloud data centers. Moreover, the study has not explored cloud-native interactive workloads and complex workflows, which are increasingly prevalent in modern cloud-computing ecosystems. Future research could explore how the proposed technique can be adapted or extended to cater to the requirements and challenges associated with cloud-native applications and intricate workflows. A heterogeneous set of accelerators, such as FPGAs, GPUs, and SmartNICs, characterizes some real-world cloud data center ecosystems. Future studies could investigate how such accelerators influence the performance of the workload management technique.

The adaption and integration of more sophisticated DRL algorithms could also be the subject of future study. Addition-ally, a thorough evaluation of DRL training using various algorithms might be carried out to find areas for optimization efficiency, sample complexity, and convergence rate improvements. This paper has mainly focused on optimizing carbon emissions and operating costs, but geo-distributed data center resource management is a multi-faceted problem. There are often conflicting goals between the cloud provider (e.g., operational efficiency, operating costs, sustainability, regulatory compliance, and maximizing uptime) and the customer (e.g., performance and reliability, billing costs, scalability, and security). Future research could explore integrating multi-objective optimization techniques within our framework that explore provider and customer objectives at the same time. Analyzing multi-objective optimization could include mechanisms to weigh and harmonize the different goals, improving overall satisfaction and performance.

## 9 CONCLUSIONS

This paper tackles the critical challenge of carbon emissions and operating costs in cloud computing by proposing an innovative optimization method for geo-distributed cloud data centers. The significant contribution lies in developing a new game-theoretic DRL, GT-DRL. Our proposed approach incorporates game-theoretic principles to achieve optimal workload distribution and resource allocation in a non-cooperative manner. The mathematical system model presented in this study encompasses various aspects, including workload arrival rates, data center execution rates, carbon emissions, renewable energy sources, data center power consumption, electricity costs, and network costs. The model extensively represents real-world factors that influence cloud data centers' efficiency and environmental impact. Unlike mathematical optimization techniques [18], [16], and [17], our approach works without any workload prediction system because of its dynamic nature.

The experimental analysis provides a comprehensive evaluation of the proposed game-theoretic DRL method as compared to recent techniques like FD [18], GA [16], NASH [17], DDPG [32], and PPO [33]. Through elaborate simulations involving diverse scenarios, such as different numbers of data centers, workload patterns, and levels of renewable energy, the proposed method consistently exhibits superior efficiency. Notably, GT-DRL is adaptable and effective in variable workload characteristics. Unlike FD [18], GA [16], and NASH [17], one of the critical strengths of the proposed method is its adeptness in utilizing renewable energy sources to a significant effect. Consequently, it intelligently reduces carbon emissions, marking it an environmentally sustainable solution. By combining the rapidness of a non-cooperative optimization strategy with the extraordinary exploration abilities of DRL, our proposed method allows for searching through a broad solution space, which makes it sample efficient and reaches the solution faster than other state-of-the-art DRL techniques. GT-DRL achieved 55%, 32%, 21%, 43%, and 5% cloud carbon emissions reductions and up to 48%, 51%, 34%, 53%, and 20% better cloud operating costs savings on average than FD [18], GA [16], NASH [17], DDPG [32], and PPO [33] respectively. In conclusion, our game-theoretic DRL algorithm, GT-DRL, significantly advances workload optimization for geo-distributed cloud data centers, allowing for faster adaption to dynamic environments, helping in avoiding local optima by considering the interplay between multiple players, and guiding exploration in DRL leading to intelligent exploration strategies. By effectively reducing carbon emissions and operating costs, it addresses economic concerns and contributes positively to environmental sustainability.

## REFERENCES

[1] "Key Internet Statistics in 2023 (Including Mobile)," [Online]. Available: https://www.broadbandsearch.net/blog/internet-statistics. [Accessed 1 June 2023].

[2] "AI and the Data Center: Challenges and Investment Strategies," [Online]. Available: https://www.informationweek.com/data-centers/ai-and-the-data-center-challenges-and-investment-strategies-. [Accessed 1 June 2023].

[3] *Cisco Global Cloud Index: Forecast and Methodology, 2016–2021,* San Jose, CA: Cisco, 2018.

[4] "Why AI inference will remain largely on the CPU," [Online]. Available: https://www.nextplatform.com/2023/04/05/why-ai-inference-will-remain-largely-on-the-cpu/. [Accessed 1 aug 2024].

[5] "AI Stocks: Why The 'Inferencing' Market Will Be Bigger Than Training Models," [Online]. Available: https://www.investors.com/news/technology/ai-stocks-market-shifting-to-inferencing-from-training/. [Accessed 01 August 2023].

[6] "Infrastructure Requirements for AI Inference vs. Training," [Online]. Available: https://www.hpcwire.com/2022/06/13/infrastructure-requirements-for-ai-inference-vs-training/. [Accessed 1 August 2024].

[7] "Unlock the Power of PowerEdge Servers for AI Workloads: Experience Up to 177% Performance Boost!," [Online]. Available: https://infohub.delltechnologies.com/en-US/p/unlock-the-power-of-poweredge-servers-for-ai-workloads-experience-up-to-177-performance-boost/. [Accessed 1 August 2024].

[8] "CPU vs GPU: How to Narrow the Deep Learning Performance Gap?," [Online]. Available: https://deci.ai/blog/narrow-the-deep-learning-performance-gap/. [Accessed 1 August 2024].

[9] "Data Center Locations," [Online]. Available: https://www.google.com/about/datacenters/locations/. [Accessed 1 May 2023].

[10] "Global Infrastructure," [Online]. Available: https://aws.amazon.com/about-aws/global-infrastructure/. [Accessed 1 May 2023].

[11] "The value of geographically diverse data centers," [Online]. Available: https://www.flexential.com/resources/blog/value-geographically-diverse-data-centers. [Accessed 1 June 2023].

[12] "What is time-of-use pricing and why is it important?," [Online]. Available: http://www.energy-exchange.net/time-of-use-pricing/. [Accessed 1 June 2017].

[13] "Dynamic pricing," [Online]. Available: http://whatis.techtarget.com/definition/dynamic-pricing. [Accessed 1 June 2017].

[14] "Demand Charges," [Online]. Available: http://www.stem.com/resources/learning/. [Accessed 1 June 2017].

[15] "Carbon emissions of data usage increasing, but what is yours?," 1 June 2023. [Online]. Available: https://www.climateneutralgroup.com/en/news/carbon-emissions-of-data-centers/.

[16] N. Hogade, S. Pasricha, H. J. Siegel, A. A. Maciejewski, M. A. Oxley and E. Jonardi, "Minimizing Energy Costs for Geographically Distributed Heterogeneous Data Centers," *IEEE Transactions on Sustainable Computing,* vol. 3, pp. 318-331, 2018.

[17] N. S. Hogade, S. Pasricha and H. J. Siegel, "Energy and Network Aware Workload Management for Geographically Distributed Data Centers," *IEEE Transactions on Sustainable Computing*, 2021.

[18] H. Goudarzi and M. Pedram, "Geographical Load Balancing for Online Service Applications in Distributed Datacenters," in *IEEE 6th Int'l Conf. on Cloud Computing (CLOUD '13)*, June 2013.

[19] V. Mnih, K. Kavukcuoglu, D. Silver, A. A. Rusu, J. Veness, M. G. Bellemare, A. Graves, M. Riedmiller, A. K. Fidjeland, G. Ostrovski, S. Petersen, C. Beattie, A. Sadik, I. Antonoglou, H. King, D. Kumaran, D. Wierstra, S. Legg and D. Hassabis, "Human-level control through deep reinforcement learning," *Nature*, vol. 518, no. 7540, pp. 529-533, 2 2015.

[20] N. Hogade and S. Pasricha, "A Survey on Machine Learning for Geo-Distributed Cloud Data Center Managements," *IEEE Transactions on Sustainable Computing*, vol. 8, no. 1, pp. 15-31, 1 2023.

[21] F. Rossi, V. Cardellini, F. L. Presti and M. Nardelli, "Geo-distributed efficient deployment of containers with Kubernetes," *Computer Communications*, vol. 159, p. 161–174, 2020.

[22] X. Zhou, K. Wang, W. Jia and M. Guo, "Reinforcement learning-based adaptive resource management of differentiated services in geo-distributed data centers," in *2017 IEEE/ACM 25th International Symposium on Quality of Service (IWQoS)*, 2017.

[23] C. Xu, K. Wang and M. Guo, "Intelligent resource management in blockchain-based cloud datacenters," *IEEE Cloud Computing*, vol. 4, p. 50–59, 2017.

[24] Q. Li, Z. Peng, D. Cui, J. He, K. Chen and J. Zhou, "Data Center Selection Based on Reinforcement Learning," in *2019 4th International Conference on Cloud Computing and Internet of Things (CCIOT)*, 2019.

[25] D. Cheng, X. Zhou, Z. Ding, Y. Wang and M. Ji, "Heterogeneity aware workload management in distributed sustainable datacenters," *IEEE Transactions on Parallel and Distributed Systems*, vol. 30, p. 375–387, 2018.

[26] T. Wei, S. Ren and Q. Zhu, "Deep reinforcement learning for joint datacenter and hvac load control in distributed mixed-use buildings," *IEEE Transactions on Sustainable Computing*, 2019.

[27] E. Baccour, A. Erbad, A. Mohamed, F. Haouari, M. Guizani and M. Hamdi, "RL-OPRA: Reinforcement Learning for Online and Proactive Resource Allocation of crowdsourced live videos," *Future Generation Computer Systems*, vol. 112, p. 982–995, 2020.

[28] C. Xu, K. Wang, P. Li, R. Xia, S. Guo and M. Guo, "Renewable energy-aware big data analytics in geo-distributed data centers with reinforcement learning," *IEEE Transactions on Network Science and Engineering*, vol. 7, p. 205–215, 2018.

[29] Y. Qin, W. Han, Y. Yang and W. Yang, "Joint energy optimization on the server and network sides for geo-distributed data centers," *The Journal of Supercomputing*, vol. 77, p. 7757–7790, 2021.

[30] H. Wang, H. Shen, Z. Li and S. Tian, "GeoCol: A Geo-distributed Cloud Storage System with Low Cost and Latency using Reinforcement Learning," in *2021 IEEE 41st International Conference on Distributed Computing Systems (ICDCS)*, 2021.

[31] J. Bi, Z. Yu and H. Yuan, "Cost-optimized Task Scheduling with Improved Deep Q-Learning in Green Data Centers," *2022 IEEE International Conference on Systems, Man, and Cybernetics (SMC)*, pp. 556-561, 10 2022.

[32] T. Tang, B. Wu and G. Hu, "A Hybrid Learning Framework for Service Function Chaining Across Geo-Distributed Data Centers," *IEEE Access*, vol. 8, p. 170225–170236, 2020.

[33] J. Zhao, M. A. Rodriguez and R. Buyya, "A Deep Reinforcement Learning Approach to Resource Management in Hybrid Clouds Harnessing Renewable Energy and Task Scheduling," *2021 IEEE 14th International Conference on Cloud Computing (CLOUD)*, pp. 240-249, 9 2021.

[34] H. Wang, H. Shen, J. Gao, K. Zheng and X. Li, "Multi-Agent Reinforcement Learning based Distributed Renewable Energy Matching for Datacenters," in *50th International Conference on Parallel Processing*, 2021.

[35] A. De Santis, T. Giovannelli, S. Lucidi, M. Messedaglia and M. Roma, "Determining the optimal piecewise constant approximation for the nonhomogeneous Poisson process rate of Emergency Department patient arrivals," *Flexible Services and Manufacturing Journal*, vol. 34, no. 4, pp. 979-1012, 12 2022.

[36] R. Traylor, *A Stochastic Reliability Model of a Server under a Random Workload*, arXiv preprint arXiv:1511.04130, 2015.

[37] D. Dauwe, E. Jonardi, R. D. Friese, S. Pasricha, A. A. Maciejewski, D. A. Bader and H. J. Siegel, "HPC node performance and energy modeling with the co-location of applications," *The J. of Supercomputing*, vol. 72, pp. 4771-4809, 01 December 2016.

[38] "AIBench: A Comprehensive AI Benchmark Suite for Datacenter, HPC, Edge, and AIoT," [Online]. Available: https://www.benchcouncil.org/aibench/. [Accessed 1 June 2023].

[39] "U.S. Electric Power Industry Estimated Emissions by State," [Online]. Available: https://www.eia.gov/electricity/data/state/emission_annual.xlsx. [Accessed 1 June 2023].

[40] Brockman Greg, Cheung Vicki, Pettersson Ludwig, Schneider Jonas, Schulman John, Tang Jie and ZarembaWojciech, "OpenAI Gym," *arXiv preprint*, 2016.

[41] S. Imambi, K. B. Prakash and G. R. Kanagachidambaresan, "PyTorch," 2021, pp. 87-104.

[42] A. Raffin, A. Hill, A. Gleave, A. Kanervisto, M. Ernestus and N. Dormann, "Stable-Baselines3: Reliable Reinforcement Learning Implementations," *J. Mach. Learn. Res.*, vol. 22, no. 1, 1 2021.

[43] D. Zhang, D. Dai, Y. He, F. S. Bao and B. Xie, "RLScheduler: An automated HPC batch job scheduler using reinforcement learning," in *SC20: International Conference for High Performance Computing, Networking, Storage and Analysis*, 2020.

[44] D. Whitley, "The GENITOR algorithm and selective pressure: Why rank-based allocation of reproductive trials is best," in *3rd International Conf. on Genetic Algorithms*, June 1989.

[45] "Amazon CloudFront Pricing," [Online]. Available: https://aws.amazon.com/cloudfront/pricing/. [Accessed 1 May 2020].

[46] NREL, "National Solar Radiation Database," [Online]. Available: https://mapsbeta.nrel.gov/nsrdb-viewer/. [Accessed 1 June 2017].

[47] "Net Metering," [Online]. Available: http://freeingthegrid.org/. [Accessed 1 June 2017].

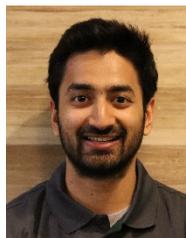

**Ninad Hogade** received his B.E. in Electronics Engineering from Vishwakarma Institute of Technology, India, and his M.S. and Ph.D. in Computer Engineering from Colorado State University, USA. He is currently a research scientist at Hewlett Packard Labs, USA. His research interests include Reinforcement Learning, scheduling in high performance and cloud computing.

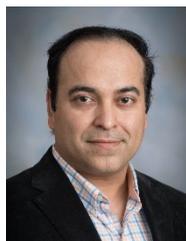

**Sudeep Pasricha** received his B.E. in Electronics and Communications from the Delhi Institute of Technology, India, and his M.S. and Ph.D. in Computer Science from the University of California, Irvine. He is currently a Professor and Chair of Computer Engineering at Colorado State University. He is a Fellow of the IEEE and an ACM Distinguished Member. Homepage: http://www.engr.colostate.edu/sudeep.